\documentclass[smallextended]{svjour2}
\usepackage{amsmath}
\newcommand{\fpoperator}{\hat{\mathcal{L}}_{\mathbf \Lambda}}
\newcommand{\greens}{g_{\mathbf\Lambda}(x,x^\prime)}
\newcommand{\ps}{P^{s}_{\mathbf\Lambda}}
\newcommand{\js}{J^{s}_{\mathbf\Lambda}}
\newcommand{\current}{\hat{\mathcal{J}}_{\mathbf\Lambda}}

\begin{document}

\title{Exact formula for currents in strongly pumped diffusive systems}

\author{Jordan M.\ Horowitz \and Christopher Jarzynski}

\institute{J.\ M.\ Horowitz \at
Department of Physics, University of Maryland, College Park, MD 20742 USA \\
Tel.: \\
Fax: \\
\email{horowitz@umd.edu}
\and
C. Jarzynski \at
Department of Chemistry and Biochemistry, and Institute for Physical Science and Technology,\\
University of Maryland, College Park, MD 20742 USA
}

\date{\today}

\maketitle

\begin{abstract}
We analyze a generic model of mesoscopic machines driven by the nonadiabatic variation of external parameters.
We derive a formula for the probability current; as a consequence we obtain a no-pumping theorem for cyclic processes satisfying detailed balance and demonstrate that the rectification of current requires broken spatial symmetry.
\end{abstract}

Molecular motors and other mesoscopic machines perform their duties in the face of the violent thermal agitation of the microscopic world~\cite{Astumian2002}.
Living organisms employ a host of such machines
to pump ions across cell membranes, transport cargo, and cause muscle contraction, among many other tasks~\cite{Howard,Kolomeisky2007}.
Remarkably, artificial nanoscale machines have in recent years been synthesized and investigated in laboratories around the world~\cite{Kay2007,Siwy2002}.
Apart from their intrinsic allure these machines suggest novel technologies such as microsopic particle segregators~\cite{Astumian1997,Hanggi2008,Faucheux1995}.
In view of the diversity of natural and artificial mesoscopic machines, there is growing interest in developing theoretical frameworks to provide insight, guide experiments, and stimulate conceptual breakthroughs.

The nanoscale machines found in biological systems are typically {\it autonomous}, powered by the consumption of chemical fuel (such as ATP) or some other steady supply of energy (e.g.\ sunlight).
By contrast, many artificial machines are {\it non-autonomous}: external parameters, such as electric fields, temperature, pressure, and chemical potentials, are varied with time to produce the desired behavior.
For such machines, a natural goal of theory is to describe how the system responds to these time-dependent perturbations.

In this article we analyze {\it stochastic pumps}, a generic model of non-autonomous mesoscopic machines operating in the presence of random noise.
We specifically focus on systems that evolve diffusively in one dimension -- Brownian motors provide a concrete example~\cite{Astumian2002,Astumian1997,Hanggi2008,Reimann2002} -- and we are interested in the current $J(x,t)$ that arises due to the ``pumping'' of external parameters ${\mathbf\Lambda}(t)$.
(Discrete-state stochastic pumps have long been studied in biological systems \cite{Tsong1988}.
For recent analyses of discrete-state stochastic pumps, see Refs.~\cite{Sinitsyn2007,Sinitsyn2007b,Astumian2007,Rahav2008,Chernyak2008}.)
Parrondo~\cite{Parrondo1998} has formulated this problem similarly, and has solved for the current induced by adiabatic (i.e.\ slow) pumping.
Here we analyze the nonadiabatic case in which the parameters are manipulated at arbitrary rates.
We derive a nontrivial expression for $J(x,t)$ (Eq.~\ref{eq:decomposition}) and we explore its consequences.
In particular, we obtain a geometric expression for the integrated current in the limit of slow pumping (Eq.~\ref{eq:adiabatic});
we extend a no-pumping theorem that we had previously obtained for discrete systems~\cite{Rahav2008};
and we provide a simple mathematical argument for the widely known fact that the generation of nonzero integrated current requires broken spatial symmetry \cite{Reimann2002,Hanggi2008}.

Let us consider a system that can be modeled as a one-dimensional diffusion process~\cite{Gardiner}  on the circle.
The state of the process at time $t$ is denoted by $x(t)$ and takes values in the range $x\in[0,L]$, with the end points identified.
This class of diffusion processes encompasses variables that are intrinsically periodic, such as the dihedral angle of a chemical bond, as well as extended reaction coordinates evolving in a periodic potential~\cite{Reimann2002}.
The  probability density $P(x,t)$ to observe state $x$ at time $t$ evolves according to the Fokker-Planck equation \cite{Risken}
\begin{equation}\label{eq:fp}
\begin{aligned}
\frac{\partial}{\partial t}P(x,t) &= \left[-\frac{\partial}{\partial x}A_{\mathbf\Lambda}(x)+\frac{\partial^2}{\partial x^2}B_{\mathbf\Lambda}(x)\right] P(x,t) \\ 
&\equiv\fpoperator(x)P(x,t),
\end{aligned}
\end{equation}
with periodic boundary conditions.
The dynamics depend on the external parameters ${\mathbf\Lambda}$ through the drift coefficient $A_{\mathbf\Lambda}(x)$ and diffusion coefficient $B_{\mathbf\Lambda}(x)$ which are assumed to be periodic in $x$.
The system is ``pumped'' by varying the parameters in time from $t=0$ to $\tau$ according to a specified protocol ${\mathbf\Lambda}(t)$, from ${\mathbf\Lambda}(0)={\mathbf a}$ to ${\mathbf\Lambda}(\tau)={\mathbf b}$, and {\it we are interested in the flow of probability that arises as a result of this pumping}.
(For brevity, the time-dependence of ${\mathbf\Lambda}$ will be left implicit throughout the following.)

Since probability is conserved, it is natural to cast Eq.~\ref{eq:fp} as a continuity equation,
\begin{equation}\label{eq:continuity}
\frac{\partial}{\partial t}P(x,t)+\frac{\partial}{\partial x}J(x,t)=0,
\end{equation}
where the {\it instantaneous current}
\begin{equation}\label{eq:current}
\begin{split}
J(x,t) &= \left[ A_{\mathbf{\Lambda}}(x) - \frac{\partial}{\partial x} B_{\mathbf\Lambda}(x) \right] P(x,t)\\
& \equiv\hat{\mathcal{J}}_{{\mathbf\Lambda}}(x)P(x,t)
\end{split}
\end{equation}
is the rate of flow of probability, in the positive direction, past a fixed observation point $x$ at time $t$.
We will also investigate the {\it integrated current}
\begin{equation}\label{eq:int}
\Phi(x) = \int_0^\tau {\mathrm d}t \, J(x,t),
\end{equation}
which measures the net flow of probability past the point $x$ during the time interval $0<t<\tau$.

To analyze such diffusion processes, it is convenient to define two {\it auxiliary functions} $\psi_{\mathbf\Lambda}(x)$ and $\varphi_{\mathbf\Lambda}(x)$, which are not necessarily periodic in $x$:
\begin{equation}\label{eq:potential}
\varphi_{\mathbf\Lambda}(x)=\ln B_{\mathbf\Lambda}(x)-\int_0^x\mathrm{d}y\frac{A_{\mathbf\Lambda}(y)}{B_{\mathbf\Lambda}(y)}\equiv\ln B_{\mathbf\Lambda}(x)+\psi_{\mathbf\Lambda}(x) .
\end{equation}
We will refer to $\varphi_{\mathbf\Lambda}(x)$ as the {\it potential}, in view of the role this function plays when the dynamics satisfy detailed balance (see below).
Observe that $A_{\mathbf\Lambda}(x)$ and $B_{\mathbf\Lambda}(x)$ can be reconstructed from $\varphi_{\mathbf\Lambda}(x)$ and $\psi_{\mathbf\Lambda}(x)$, that is Eq.~\ref{eq:potential} can be inverted.
In other words the diffusion process is characterized equally well by the auxiliary functions as by the drift and diffusion coefficients.

For every fixed ${\mathbf\Lambda}$, there exists a unique stationary distribution $\ps(x)$, satisfying $\fpoperator\ps=0$, with stationary current $\js=\current(x)\ps(x)$
\footnote{ From Eq.~\ref{eq:continuity} it follows that $\js$ does not depend on $x$: in the stationary state, the same current flows past every observation point $x$.}.
In the special case that $\varphi_{\mathbf\Lambda}(0)=\varphi_{\mathbf\Lambda}(L)$, it follows from Eqs.~\ref{eq:fp} and \ref{eq:current} that $\ps(x)\propto e^{-\varphi_{\mathbf\Lambda}(x)}$ and $\js=0$.
In this case, we will say that $\fpoperator$ satisfies detailed balance and we will view the stationary distribution as an equilibrium distribution, $\ps=P^{eq}_{\mathbf\Lambda}$.

For the setup we have just outlined, there are two distinct mechanisms to produce currents~\cite{Chernyak2006}.
First, when detailed balance is broken the stationary distributions themselves support non-zero currents, $\js\ne 0$.
Second, if the system is driven away from the stationary distribution by varying ${\mathbf\Lambda}$ with time, currents arise due to the resulting redistribution of probability.
In this article we derive an explicit decomposition of the total current into these two contributions:
\begin{equation}\label{eq:decomposition}
J(x,t)=J^{s}_{{\mathbf \Lambda}}+\int_0^L\mathrm{d}x^\prime \mathcal{V}_{{\mathbf \Lambda}}(x,x^\prime)\dot P(x^\prime,t),
\end{equation}
where an analytic expression for the integral kernel $\mathcal{V}_{\mathbf\Lambda}(x,x^\prime)$ is given in Eq.\ \ref{eq:v} below.
This {\it exact} result gives the net current as the sum of a baseline stationary contribution $\js$ and an excess or ``pumped'' contribution $J^{ex}(x,t)$ associated with the variation of external parameters.
This decomposition is analogous to that of heat flow encountered in steady-state thermodynamics \cite{Oono1997,Hatano2001}.
As we will show, Eq.~\ref{eq:decomposition} is amenable to further analysis, and provides a useful theoretical tool for studying the response of continuous stochastic pumps under arbitrary driving conditions.

Note that $\mathcal{V}_{\mathbf\Lambda}(x,x^\prime)$ is not defined uniquely: 
Eq.~\ref{eq:decomposition} is unaffected by a transformation of the form $\mathcal{V}_{\mathbf\Lambda}(x,x^\prime) \rightarrow \mathcal{V}_{\mathbf\Lambda}(x,x^\prime) + f(x)$, since probability is conserved.

To derive Eq.\ \ref{eq:decomposition}, we first solve for $P(x,t)$ in terms of $\dot P(x,t)$ (Eq.~\ref{eq:solution}), and then combine that result with Eq.\ \ref{eq:current} to determine $J(x,t)$.
To this end, let us take the following atypical view of the Fokker-Planck equation: for fixed $t$ let us interpret Eq.\ \ref{eq:fp} as an operator equation for $P(x,t)$  with operator $\fpoperator(x)$ and {\it source} term $\dot P(x,t)$. 
Ordinarily, we solve an operator equation by finding the inverse operator.
However, since our Fokker-Planck operator has a null eigenvector, $\fpoperator\ps=0$, it is not invertible.
Therefore, we instead introduce the integral operator $\hat{\mathcal{R}}_{\mathbf\Lambda}(x)=\int\mathrm{d}x^\prime\greens$ which is the {\it pseudoinverse} of $\fpoperator(x)$ (see below for a brief definition of pseudoinverse in this context).  Here, the integral kernel $\greens$ is the {\it modified Green's function} for $\fpoperator(x)$ \cite{Stakgold}, defined as the solution of the boundary value problem
\begin{equation}\label{eq:greens}
\left\{
\begin{aligned}
&\fpoperator^\dag(x^\prime)\greens=\delta(x^\prime-x)-\ps(x) \\
&\greens\Big\vert_{x^\prime=0}^{x^\prime=L}=0, \qquad \frac{\partial \greens}{\partial x^\prime}\Big\vert_{x^\prime=0}^{x^\prime=L}=0
\end{aligned}
\right .,
\end{equation}
where $\fpoperator^\dag(x)=A_{\mathbf\Lambda}(x)\partial/\partial x+B_{\mathbf\Lambda}(x)\partial^2/\partial x^2$ is the formal adjoint of $\fpoperator(x)$.
The term $\ps(x)$ in Eq.~\ref{eq:greens} accounts for the fact that $\fpoperator(x)$ is not invertible; without this term the boundary value problem has no solution~\cite{Stakgold}.
Since Eq.~\ref{eq:greens} is unaffected by a replacement $\greens \rightarrow \greens + f(x)$, the solution of Eq.\ \ref{eq:greens} is not unique.
This is the source of the non-uniqueness of $\mathcal{V}_{\mathbf\Lambda}(x,x^\prime)$ mentioned above.

As mentioned, $\hat{\mathcal{R}}_{\mathbf\Lambda}$ is the pseudoinverse of $\fpoperator$.
That is, in place of the usual inverse property ($\hat{\mathcal{R}}_{\mathbf\Lambda}\fpoperator=\hat{\mathcal{I}}$), $\hat{\mathcal{R}}_{\mathbf\Lambda}$ satisfies the inverse-like property
\begin{equation}\label{eq:inverse}
\begin{aligned}
\hat{\mathcal{R}}_{\mathbf\Lambda}\fpoperator P(x,t)
&\equiv \int_0^L\mathrm{d}x^\prime\greens\fpoperator(x^\prime)P(x^\prime,t) \\ 
&= P(x,t)-\ps(x),
\end{aligned}
\end{equation}
where we have twice integrated by parts and exploited Eq.\ \ref{eq:greens}.
We see that $\hat{\mathcal{R}}_{\mathbf\Lambda} \fpoperator$ projects onto the orthogonal complement of the null space of $\fpoperator$~\cite{Cardus}.
Simply put, $\hat{\mathcal{R}}_{\mathbf\Lambda}$  acts as an inverse on the subspace where $\fpoperator$ is invertible.

We now apply $\hat{\mathcal{R}}_{\mathbf\Lambda}$ to both sides of Eq.\ \ref{eq:fp}, then use the pseudoinverse property (Eq.\ \ref{eq:inverse}) to obtain
\begin{equation}\label{eq:solution}
P(x,t)=P^{s}_{{\mathbf\Lambda}}(x)+\int_0^L\mathrm{d}x^\prime g_{{\mathbf\Lambda}}(x,x^\prime)\dot P(x^\prime,t).
\end{equation}
Next we apply the current operator (Eq.~\ref{eq:current}) to both sides of this equation.
This gives us
\begin{equation}
J(x,t) = J^{s}_{{\mathbf\Lambda}}+\int_0^L\mathrm{d}x^\prime\hat{\mathcal{J}}_{{\mathbf\Lambda}}(x)g_{{\mathbf\Lambda}}(x,x^\prime)\dot P(x^\prime,t).
\end{equation}
Comparing with Eq.\ \ref{eq:decomposition} we see that 
$\mathcal{V}_{\mathbf\Lambda}=\current g_{\mathbf\Lambda}$.
Finally, we apply $\current(x)$ to Eq.~\ref{eq:greens} to arrive at
\begin{equation}\label{eq:diffeqV}
\left\{
\begin{aligned}
&\fpoperator^\dag(x^\prime)\mathcal{V}_{\mathbf\Lambda}(x,x^\prime)=\current(x)\delta(x^\prime-x)-\js \\
&\mathcal{V}_{\mathbf\Lambda}(x,x^\prime)\Big\vert_{x^\prime=0}^{x^\prime=L}=0, 
\qquad \frac{\partial \mathcal{V}_{\mathbf\Lambda}(x,x^\prime)}{\partial x^\prime}\Big\vert_{x^\prime=0}^{x^\prime=L}=0
\end{aligned}
\right ..
\end{equation}
This boundary value problems is solved by (see Appendix \ref{sec:appendix})
\begin{equation}\label{eq:v}
\mathcal{V}_{\mathbf\Lambda}(x,x^\prime) = \left(1+J^{s}_{\mathbf{\Lambda}}\tau_{\mathbf\Lambda}(L)\right)
\pi_{\mathbf\Lambda}(x^\prime) +\theta(x^\prime-x)+J^{s}_{\mathbf{\Lambda}}\tau_{\mathbf\Lambda}(x^\prime),
\end{equation}
where $\theta(x^\prime-x)$ is the Heaviside step function;
$\varphi_{\mathbf\Lambda}$ and $\psi_{\mathbf\Lambda}$ are given in Eq.\ \ref{eq:potential}; and we have introduced the {\em splitting probability}
\begin{equation}\label{eq:split}
\pi_{\mathbf\Lambda}(x)=\frac{\int_x^L\mathrm{d}ye^{\psi_{\mathbf\Lambda}(y)}}{\int_0^L\mathrm{d}ye^{\psi_{ {\mathbf{\Lambda}}}(y)}} ,
\end{equation}
and the \emph{conditional mean first exit time}
\begin{equation}\label{eq:cmfet}
\tau_{\mathbf\Lambda}(x)=\int_0^{x}\mathrm{d}y\int_0^y\mathrm{d}ze^{\psi_{\mathbf{\Lambda}}(y)-\varphi_{\mathbf{\Lambda}}(z)}.
\end{equation}
The splitting probability and the conditional mean first exit time have the following interpretations~\cite{Gardiner}:
with $\mathbf\Lambda$ fixed, if the system evolves from $x_0$ until it first exits the domain $[0,L]$, then $\pi_{\mathbf\Lambda}(x_0)$ is the probability that this exit will occur at $x=0$, rather than $x=L$; and $\tau_{\mathbf\Lambda}(x_0)$ is the average time for the system to make this first exit through $x=0$.
Roughly speaking, the splitting probability measures the relative likelihood for the process to go clockwise versus counterclockwise around the circle.

Equations \ref{eq:decomposition} and \ref{eq:v} specify the current $J(x,t)$, in terms of the rate of change of the probability distribution $\dot P(x^\prime,t)$.
We now investigate consequences of this result.

Let us define the {\it excess integrated current} $\Phi^{ex}(x)$ to be the net current pumped across the point $x$, in excess of the time-integrated, baseline stationary flow, $\Phi^{s}=\int\mathrm{d}tJ^{s}_{{\mathbf\Lambda}(t)}$.
By Eq.~\ref{eq:decomposition},
\begin{equation}\label{eq:excess}
\Phi^{ex}(x)=\int_0^\tau\mathrm{d}t\int_0^L\mathrm{d}x^\prime\mathcal{V}_{{\mathbf\Lambda}}(x,x^\prime)\dot P(x^\prime,t) .
\end{equation}
If ${\mathbf\Lambda}$ is varied very slowly from ${\mathbf a}$ to ${\mathbf b}$, the system remains near the stationary distribution, $P(x,t)\sim P^{s}_{{\mathbf\Lambda}(t)}(x)$.
This suggests that in the adiabatic limit we can make the substitution
$\dot P(x,t)\mathrm{d}t \to {\mathbf\nabla}_{\mathbf\Lambda}\ps(x) \cdot \mathrm{d}{\mathbf\Lambda}$ \footnote{This substitution can be justified by appealing to an adiabatic perturbation theory. J. Horowitz, S. Vaikuntanathan, S. Rahav, and C. Jarzynski (unpublished).} in Eq.\ \ref{eq:excess} to find
\begin{equation}\label{eq:adiabatic}
\Phi^{ex}(x)=\int_{\mathbf a}^{\mathbf b}\mathrm{d}{\mathbf\Lambda}\cdot{\mathbf\Sigma}_{\mathbf\Lambda}(x),
\end{equation}
where ${\mathbf\Sigma}_{\mathbf\Lambda}(x)=\int\mathrm{d}x^\prime\mathcal{V}_{\mathbf\Lambda}(x,x^\prime)\mathbf{\nabla}_{\mathbf\Lambda} P^s_{\mathbf\Lambda}(x^\prime)$.
This result is {\it geometric}: time no longer explicitly appears, and the excess current is determined solely by the path taken from ${\mathbf a}$ to ${\mathbf b}$ in parameter space.
If the drift and diffusion coefficients take the special form $A_{\mathbf\Lambda}(x) = (\partial/\partial x) V_{\mathbf\Lambda}(x)$ and $B_{\mathbf\Lambda}(x)=D$, 
Eq.\ \ref{eq:adiabatic} reduces to a result obtained by Parrondo~\cite{Parrondo1998}.
Analogous geometric expressions for the adiabatic integrated current have also been derived for discrete stochastic pumps \cite{Rahav2008,Sinitsyn2007,Sinitsyn2007b,Astumian2007}.

In Ref.~\cite{Rahav2008} a ``no-pumping'' theorem was derived for discrete stochastic pumps.
Briefly, this result stated that in order to obtain non-zero integrated currents for cyclic processes, it is necessary to vary both the energy levels of the discrete states and the effective energy barriers associated with transitions between these states.
Subsequently, Chernyak and Sinitsyn have pointed to the possibility that a similar result applies to continuous stochastic pumps  \cite{Chernyak2008}.
Here we explicitly develop a no-pumping theorem for continuous stochastic pumps.

Within the general model analyzed above, let us now restrict ourselves to the case that detailed balance holds for all ${\mathbf\Lambda}$, hence $\js=0$.
Imagine that the parameters are varied periodically with time, from the distant past, with period $\tau$, so that by $t=0$ the system has settled into a periodic steady state:
$P(x,t) = P(x,t+\tau)$~\footnote{
As in Ref.~\cite{Rahav2008}, our analysis also applies if we begin in the equilibrium distribution $P_{\mathbf\Lambda}^{\rm eq}$, then vary ${\mathbf\Lambda}$ at arbitrary rate around a closed loop in parameter space, then allow the system to relax back to equilibrium.}.
In this setup, which is regularly encountered in the theory of Brownian motors \cite{Astumian2002,Astumian1997,Hanggi2008,Reimann2002},
the current $J(x,t)$ at a fixed location $x$ evolves periodically with time.
It is then natural to consider the integrated current over one cycle, which for a cyclic process with detailed balance ($J^s_{\bf \Lambda}=0$) is
\begin{equation}\label{eq:integrated}
\Phi=\int_0^\tau\mathrm{d}t\int_0^L\mathrm{d}x^\prime\pi_{{\mathbf\Lambda}}(x^\prime)\dot P(x^\prime,t);
\end{equation}
see Eqs.\ \ref{eq:int}, \ref{eq:decomposition}, and \ref{eq:v}.
The value of $\Phi$ represents the net circulation of probability during one cycle.
If the probability merely sloshes back and forth without any accumulation of current, then $\Phi=0$; whereas, if $\Phi > 0$ ($\Phi<0$) then there is a nonzero flow of probability in the counterclockwise (clockwise) direction. 

We now argue that to obtain $\Phi\neq 0$ {\it both the potential $\varphi_{\mathbf\Lambda}(x)$ and the splitting probability $\pi_{\mathbf\Lambda}(x)$ must be varied during the process}.
The first of these conditions is easy to understand:
if the potential remains fixed during the process, i.e.\ $\varphi_{{\mathbf\Lambda}(t)}(x)=\varphi_{\mathbf a}(x)$, then the system simply remains in the initial equilibrium state, $P(x,t) \propto \exp[-\varphi_{\mathbf a}(x)]$, producing no currents whatsoever;
this is the ``no-go theorem'' of Ref.~\cite{Reimann2002}, section 6.4.1.
To see that the splitting probability must also be varied to produce integrated current, suppose we fix $\pi_{{\mathbf\Lambda}(t)}(x)=\pi_{\mathbf a}(x)$ but vary $\varphi_{{\mathbf\Lambda}(t)}(x)$.
Then
\begin{equation}
\Phi=\int_0^L\mathrm{d}x^\prime\pi_{\mathbf a}(x^\prime)\int_0^\tau\mathrm{d}t \, \dot P(x^\prime,t)=0,
\end{equation}
since the process is cyclic.
We can construct a heuristic interpretation of this result by recalling that $\pi_{\mathbf\Lambda}(x)$ measures the likelihood to generate clockwise rather than counterclockwise flow, as discussed following Eq.~\ref{eq:split}.
The integrand $\pi_{\mathbf\Lambda}(x^\prime)\dot P(x^\prime,t)$ appearing in Eq.~\ref{eq:integrated} then represents, roughly, the contribution to clockwise current induced by the redistribution of probability that occurs at location $x^\prime$ and time $t$, and the integrated current $\Phi$ is a sum of such contributions.
For a cyclic process, any probability that leaves the location $x^\prime$ must eventually return, thus if $\pi_{\mathbf\Lambda}(x^\prime)$ remains constant the clockwise and counterclockwise contributions ultimately cancel one another ($\Phi=0$).
If the splitting probability varies with time, then there is no reason to expect such cancellation.

This no-pumping theorem provides a concrete mathematical criterion for the generation of zero integrated current.
Actually realizing the independent variation of both the potential and the splitting probability in any particular system will greatly depend on the system's properties and may not be feasible; yet, in systems where one may locally vary the drift and diffusion coefficients independently, one may also vary the potential and splitting probability independently, as can be seen from Eqs.\ \ref{eq:potential} and \ref{eq:split}.

From Eqs.\ \ref{eq:potential} and \ref{eq:split}, we see that $\pi_{{\mathbf\Lambda}(t)}(x)$ [or equivalently $\psi_{{\mathbf\Lambda}(t)}(x)$] remains constant when the ratio of the drift to the diffusion coefficient does not depend on ${\mathbf\Lambda}$:
\begin{equation}\label{eq:nopump}
\frac{A_{\mathbf\Lambda}(x)}{B_{\mathbf\Lambda}(x)}=\Xi(x) .
\end{equation}
Thus our no-pumping theorem states that $\Phi=0$ if either (i) the potential is held fixed, or (ii) the drift and diffusion coefficients are related by Eq.\ \ref{eq:nopump}.

Finally, we show that Eq.~\ref{eq:decomposition} reproduces the known fact that the rectification of current in a periodically driven Brownian motor requires broken spatial symmetry \cite{Reimann2002,Hanggi2008}.
Specifically, we will show that when the driving protocol is time-periodic and the drift and diffusion coefficients have specific spatial symmetries, the integrated current over one period of driving is zero.
We will say that a periodic function $f(x)$ is symmetric or has even symmetry if $f(\delta+x)=f(\delta-x)$, or has odd symmetry if $f(\delta+x)=-f(\delta-x)$, for some fixed value $\delta$.
Without loss of generality we set $\delta=0$, since by a suitable coordinate shift $\delta$ can take any value.
We now assume that the drift coefficient has odd symmetry, $A_{\mathbf\Lambda}(x)=-A_{\mathbf\Lambda}(-x)$, and the diffusion coefficient has even symmetry, $B_{\mathbf\Lambda}(x)=B_{\mathbf\Lambda}(-x)$, for every ${\mathbf\Lambda}$.
These assumptions imply that the system satisfies detailed balance
and that both $\psi_{\mathbf\Lambda}(x)$ and the periodic steady state $P(x,t)=P(x,t+\tau)$ are symmetric.
Equations \ref{eq:split} and \ref{eq:integrated} then give
\begin{equation}
\Phi  = \int_0^\tau\mathrm{d}t\int_0^L\mathrm{d}x\int_0^L\mathrm{d}y\theta(y-x)\frac{e^{\psi_{\mathbf{\Lambda}}(y)}}{\int_0^L\mathrm{d}ye^{\psi_{\mathbf{\Lambda}}(y)}}\dot P(x,t).
\end{equation}
Changing variables $y\to L-y$ and $x\to L-x$, and exploiting the symmetry and periodicity of $\psi_{\mathbf\Lambda}(y)$ and $P(x,t)$, we find
\begin{equation}
\label{eq:symmetry}
\Phi = \int_0^\tau\mathrm{d}t\int_0^L\mathrm{d}x\int_0^L\mathrm{d}y\theta(x-y)\frac{e^{\psi_{\mathbf{\Lambda}}(y)}}{\int_0^L\mathrm{d}ye^{\psi_{\mathbf{\Lambda}}(y)}}\dot P(x,t) .
\end{equation}
If we now use the identity $\theta(x-y)=1-\theta(y-x)$ and invoke conservation of normalization, $\int_0^L\mathrm{d}x\, \dot P(x,t)=0$, Eq.~\ref{eq:symmetry} becomes $\Phi = - \Phi$.
The anticipated conclusion $\Phi=0$ is then obvious.

To summarize, Eq.~\ref{eq:decomposition} provides a decomposition of the current into a stationary contribution, and an excess, ``pumped'' contribution associated with the variation of external parameters.
Using this exact result we have established necessary conditions for current generation in cyclic processes -- the potential $\varphi_{\mathbf\Lambda}(x)$ and the splitting probability $\pi_{\mathbf\Lambda}(x)$ must be varied; demonstrated that adiabatic pumping is a geometric effect (Eq.~\ref{eq:adiabatic}); and verified that current rectification requires broken spatial symmetry.
Ref.~\cite{Chernyak2008} suggests the possibility of extending these results to higher dimensions, but this remains to be done.
It would also be interesting to investigate whether the no-pumping theorem can be extended to systems without detailed balance.
We expect that the current decomposition formula will provide a useful tool for gaining a deeper understanding of the behavior of stochastic pumps.

\begin{acknowledgements}
We would like to thank N.\ Sinitsyn, D.\ Cohen, and M.\ E.\ Fisher for insightful discussions; as well as S.\ Vaikuntanathan and A.\ Ballard for a critical reading of this document.
We also gratefully appreciate the financial support of the University of Maryland.
\end{acknowledgements}

\appendix
\section{Appendix}\label{sec:appendix}
The solution to Eq.\ \ref{eq:diffeqV} is obtained by combining the homogeneous solution with the inhomogeneous solution and then applying the boundary conditions.
The two homogeneous solutions are the splitting probability which is defined by the boundary value problem \cite{VanKampen}
\begin{equation}\label{eq:diffeqSP}
\left\{
\begin{aligned}
&\fpoperator^{\dag}(x^\prime)\pi_{\mathbf\Lambda}(x^\prime)=0 \\
&\pi_{\mathbf\Lambda}(0)=1, \qquad \pi_{\mathbf\Lambda}(L)=0
\end{aligned}
\right.
\end{equation}
and any arbitrary function of $x$ alone, say $f(x)$.
The two contributions to the inhomogeneous solution, $\js\tau_{\mathbf\Lambda}(x^\prime)$ and $\theta(x^\prime-x)$, are obtained by noting that the defining boundary value problem for the conditional mean first exit time is \cite{VanKampen}
\begin{equation}\label{eq:diffeqET}
\left\{
\begin{aligned}
&\fpoperator^{\dag}(x^\prime)\ \tau_{\mathbf\Lambda}(x^\prime)= -1 \\
&\tau_{\mathbf\Lambda}(0)=0, \qquad \frac{\partial\tau_{\mathbf\Lambda}(L)}{\partial x^\prime}=0
\end{aligned}
\right.
\end{equation}
and that
\begin{eqnarray}
\fpoperator^{\dag}(x^\prime)\theta(x^\prime-x) 
&=& \left[A_{\mathbf\Lambda}(x^\prime)\frac{\partial}{\partial x^\prime}+B_{\mathbf\Lambda}(x^\prime)\frac{\partial^2}{\partial x^{\prime 2}}\right]\theta(x^\prime-x) \\
&=& \left[A_{\mathbf\Lambda}(x^\prime)+B_{\mathbf\Lambda}(x^\prime)\frac{\partial}{\partial x^{\prime}}\right]\delta(x^\prime-x) \\
&=& \left[A_{\mathbf\Lambda}(x)-\frac{\partial}{\partial x}B_{\mathbf\Lambda}(x)\right]\delta(x^\prime-x) \\
&=& \current(x)\delta(x^\prime-x).
\end{eqnarray}
Thus, the most general solution to Eq.\ \ref{eq:diffeqV} is
\begin{equation}
\mathcal{V}_{\mathbf\Lambda}(x,x^\prime)=C\pi_{\mathbf\Lambda}(x^\prime)+f(x) + \theta(x^\prime-x)+\js\tau_{\mathbf\Lambda}(x^\prime),
\end{equation}
where $C$ is an arbitrary constant.
The value of $C$ is fixed by satisfying the first boundary condition in Eq.\ \ref{eq:diffeqV}.
The second boundary condition is then automatically satisfied due to the structure of Eq.\ \ref{eq:diffeqV}.
Lastly,  we arrive at the solution in Eq.\ \ref{eq:v} by setting $f(x)=0$, which we are free to do since the solution to Eq.\ \ref{eq:diffeqV} is not unique.

To complete the solution we must solve Eqs.\ \ref{eq:diffeqSP} and \ref{eq:diffeqET}.
Both equations can be made integrable by multiplying them by the integrating factor $e^{-\varphi_{\mathbf\Lambda}(x)}$.
For example, consider Eq.\ \ref{eq:diffeqSP} for the splitting probability:
\begin{eqnarray}
&e^{-\varphi_{\mathbf\Lambda}(x)}\fpoperator^{\dag}(x)\pi_{\mathbf\Lambda}(x) = 0& \\
&e^{-\varphi_{\mathbf\Lambda}(x)}\left[A_{\mathbf\Lambda}(x)\frac{\partial}{\partial x}+B_{\mathbf\Lambda}(x)\frac{\partial^2}{\partial x^2}\right]\pi_{\mathbf\Lambda}(x) = 0& \\
\label{eq:step}
& \left[-\frac{\partial\psi_{\mathbf\Lambda}(x)}{\partial x}e^{-\psi_{\mathbf\Lambda}(x)}\frac{\partial}{\partial x}+e^{-\psi_{\mathbf\Lambda}(x)}\frac{\partial^2}{\partial x^2}\right]\pi_{\mathbf\Lambda}(x)=0& \\
& \frac{\partial}{\partial x}e^{-\psi_{\mathbf\Lambda}(x)}\frac{\partial}{\partial x}\pi_{\mathbf\Lambda}(x) = 0, & \label{eq:diffeqSP2}
\end{eqnarray}
where to get Eq.\ \ref{eq:step} we used the definitions of $\varphi_{\mathbf\Lambda}$ and $\psi_{\mathbf\Lambda}$ in Eq.\ \ref{eq:potential}.
By integrating Eq.\ \ref{eq:diffeqSP2} twice we find the solution in Eq.\ \ref{eq:split}.
A similar calculation leads to the solution in Eq.\ \ref{eq:cmfet} for the conditional mean first exit time.

\bibliographystyle{spmpsci} 

\end{document}